\documentclass[preprint,tightenlines,showpacs,showkeys,floatfix,nofootinbib,
superscriptaddress,fleqn]{revtex4}
\usepackage{graphicx}
\usepackage{graphicx}
\usepackage{bm}
\usepackage{dcolumn}
\usepackage{amsmath}
\begin{document}

\preprint{PNU-NTG-03/2005}
\preprint{RUB-TP2-01/05}
\preprint{TPJU-1/2005}
\title{Exotic and nonexotic magnetic transitions \\
in the context of the SELEX and GRAAL experiments}
\author{Hyun-Chul Kim}
\email{hchkim@pusan.ac.kr}
\affiliation{Department of Physics, and Nuclear Physics \& Radiation Technology Institute
(NuRI), Pusan National University, 609-735 Busan, Republic of Korea}
\author{Maxim Polyakov}
\email{Maxim.Polyakov@tp2.ruhr-uni-bochum.de}
\affiliation{Institut f\"ur Theoretische Physik II, Ruhr-Universit\" at Bochum, D--44780
Bochum, Germany}
\affiliation{Petersburg Nuclear Physics Institute, Gatchina, St. Petersburg 188300, Russia}
\author{Micha{\l} Prasza{\l}owicz}
\email{michal@th.if.uj.edu.pl}
\affiliation{M. Smoluchowski Institute of Physics, Jagellonian University, ul. Reymonta
4, 30-059 Krak{\'o}w, Poland}
\author{Ghil-Seok Yang}
\email{gsyang@pusan.ac.kr}
\affiliation{Department of Physics, and Nuclear Physics \& Radiation Technology Institute
(NuRI), Pusan National University, 609-735 Busan, Republic of Korea}
\author{Klaus Goeke}
\email{Klaus.Goeke@tp2.ruhr-uni-bochum.de}
\affiliation{Institut f\"ur Theoretische Physik II, Ruhr-Universit\" at Bochum, D--44780
Bochum, Germany}

\begin{abstract}
We calculate magnetic transition moments in the chiral quark-soliton model,
with explicit SU(3)-symmetry breaking taken into account. The dynamical
model parameters are fixed by experimental data for the magnetic moments of
the baryon octet and from the recent measurements of $\Theta^{+}$ mass.
Known magnetic transition moments $\mu_{\Lambda\Sigma}$, $\mu_{N\Delta}$ are
reproduced and predictions for other octet-decuplet and octet-antidecuplet
transitions are given. In particular $\mu_{\Sigma\Sigma^{\ast}}$ recently
constrained by SELEX is shown to be below $0.82\,\mu_N$. The recent GRAAL
data on $\eta$ photoproduction off the nucleon are explained in terms of a
new narrow antidecuplet neutron-like resonance.
\end{abstract}

\maketitle

\section{Introduction}

Recent discovery of the exotic $\Theta^{+}$ pentaquark state ($uudd\bar{s}$)
by the LEPS collaboration~\cite{Nakano:2003qx} and its further confirmation
by a number of other experiments \cite{experiments}, together with still
unconfirmed observation of an exotic $\Xi_{\overline{10}}$ states by the
NA49 experiment at CERN~\cite{Alt:2003vb} renewed interest in baryon
spectroscopy. Experimental searches for these new states were motivated by
the theoretical prediction of the chiral quark-soliton model~\cite%
{Diakonov:1997mm}, where masses and decay widths of exotic antidecuplet
baryons were predicted. In fact, exotic SU(3) representations containing
exotic baryonic states are naturally accommodated within the chiral soliton
models~\cite{su3quant,Chemano} and early estimates of their masses are given
in Refs.\cite{Biedotha,Praszalowicz:2003ik}.

The findings of the pentaquark baryon $\Theta ^{+}$ and possibly of $\Xi _{%
\overline{10}}$ have triggered intensive theoretical investigations which
are summarized in Refs.\cite{Jennings:2003wz,Zhu:2004xa}. Despite
differences in theoretical models of pentaquarks it is by most theorists
agree upon the fact that $\Theta ^{+}$ is the $Y=2$, $I=0$ member of the
antidecuplet representation of SU(3) flavor and $\Xi _{\overline{10}}$ has $%
Y=-1$ and $I=3/2$ and the lowest possible spin is $1/2$. Models differ as
far as parity is concerned, where the chiral quark-soliton model ($\chi $%
QSM) considered here predicts positive parity. If so, cryptoexotic
nucleon-like and $\Sigma $-like pentaquark states that span the remainder of 
$\overline{10}$ should also exist. These states most probably will mix with
the nearby nonexotic excited three quark states and with higher exotic
states like the members of flavor $27$-plet for example. It is of utmost
importance to decide whether these cryptoexotic states are the known
resonances of the particle data group (PDG) that were misinterpreted as
three quark states or if there exist new, possibly narrow resonances that
eluded detection so far.  A recent modified partial wave analysis of $\pi N$
scattering \cite{Arndt:2003ga}, a new GRAAL experiment on $\eta $
photoproduction off the neutron \cite{GRAAL}, and the preliminary STAR
data~\cite{STAR} suggest that such new nucleon-like states may indeed
exist.  In particular, the GRAAL data are consistent with the antidecuplet
interpretation~\cite{Polyakov:2003dx} of the narrow peak around the mass
1670 MeV.

In the present work, we address the question of magnetic transitions which
are relevant for the interpretation of the GRAAL data. Recently, four of the
present authors calculated the magnetic moments of the exotic and
cryptoexotic pentaquarks, within the framework of the $\chi$QSM
including chiral symmetry breaking effects due to the nonzero strange 
quark mass in the so called \emph{model-independent approach}~\cite{Yang}.

Such an approach -- introduced to our knowledge for the first time by Adkins
and Nappi~\cite{AdNap} in the context of the Skyrme model -- can be viewed
from two perspectives. Firstly, it can be considered as a QCD-motivated tool
to analyze and classify (in terms of powers of $m_{\mathrm{s}}$ and
$1/N_{\mathrm{c}}$) the symmetry breaking terms for a given
observable. For nontrivial operators such as magnetic moments or axial
form factors a general analysis, without refering to some specific
model, is often virtually impossible. Secondly, it also provides
information for the model builders. It tells us what are the best
predictions the model can ever produce. Indeed, model calculations are
not as unique as one might think: They depend on adopted
regularization, cutoff parameters, constituent quark mass, etc. On the
other hand the success of such an analysis gives a strong hint for the
model builders that the model is correct and worth exploring.  In fact,
this concerns all the hedgehog models which would give the 
collective structure identical to the one of the $\chi$QSM including the
Skyrme model.  However, for the simplest version of the Skyrme model, some
of the parameters multiplying the group theoretical operators are
identically zero, whereas in the $\chi$QSM they do not vanish due to the
presence of the valence level.

The discovery of $\Theta ^{+}$ and possibly of $\Xi _{\overline{10}}$
constrained certain parameters of the $\chi$QSM that were previously
undetermined.  This new phenomenological input allowed us to revise previous
estimates of magnetic moments both for nonexotic
~\cite{Kim:1997ip,Kim:1998gt} and exotic baryons \cite{Kim:2003ay}. In
particular, it turned out that magnetic moment of $\Theta ^{+}$ is
negative and rather sensitive to the residual freedom which we
parameterize in terms of the pion nucleon sigma term: $\Sigma _{\pi N}$.

In this paper, we extend the recent analysis of Ref.\cite{Yang} from
magnetic moments to magnetic transitions.  Using the parametrization obtained
in Refs.\cite{Yang,Praszalowicz:2004dn}, we show that the $\chi$QSM
reproduces well experimental and empirical data for the magnetic transitions 
$\mu_{\Lambda^0\Sigma^0}$ and $\mu_{N\Delta}$ and are rather
insensitive to $\Sigma_{\pi N}$. We also calculate
$\mu_{\Sigma^-\Sigma^{\ast-}}$ and show that it is consistent with a
constraint obtained by the recent SELEX
experiment~\cite{Molchanov:2004iq}. Finally we make predictions for
$\mu_{nn_{\overline{10}}^{\ast}}$ and $\mu_{pp_{\overline
{10}}^{\ast}}$, where $n_{\overline{10}}^{\ast}$ and
$p_{\overline{10}}^{\ast}$ are nucleon-like antidecuplet states. In
particular, we show that $\mu_{pp_{\overline{10}}^{\ast}}$ is small
and almost independent of $\Sigma_{\pi N}$.  On the contrary,
$\mu_{nn_{\overline{10}}^{\ast}}$ is large and depends rather strongly
on $\Sigma_{\pi N}$. This feature allows us to explain the 
recent GRAAL data on $\eta$ production off the nucleon.

In the present work, we give explicit formulae for the magnetic transition
moments and discuss the corresponding numerical results and their
phenomenological consequences, in particular, in the context of the recent
SELEX~\cite{Molchanov:2004iq} and GRAAL experiments~\cite{GRAAL}. 

\section{General formalism}

We first recapitulate very briefly the formalism of Ref.\cite{Yang}, where
the details may be found.  The collective Hamiltonian describing baryons in
the SU(3) $\chi$QSM takes the following form \cite{Blotz:1992pw}: 
\begin{equation}
\hat{H}=\mathcal{M}_{sol}+\frac{J(J+1)}{2I_{1}}+\frac{C_{2}(\text{SU(3)}
)-J(J+1)-\frac{N_{c}^{2}}{12}}{2I_{2}}+\hat{H}^{\prime}
\end{equation}
with the symmetry breaking piece given by: 
\begin{equation}
\hat{H}^{\prime}=\alpha D_{88}^{(8)}+\beta Y+\frac{\gamma}{\sqrt{3}}
D_{8i}^{(8)}\hat{J}_{i},  \label{Hsplit}
\end{equation}
where parameters $\alpha$, $\beta$ and $\gamma$ are proportional to the
strange current quark mass $m_{s}$. Here $D_{ab}^{(\mathcal{R})}(R)$ denote
SU(3) Wigner rotation matrices and $\hat{J}$ is a collective spin operator.

Taking into account the recent experimental observation of the mass of
$\Theta^{+}$ the parameters entering Eq.(\ref{Hsplit}) can be
conveniently parameterized in terms of the pion-nucleon $\Sigma_{\pi
  N}$ term (assuming $m_{s}/(m_{u} +m_{d})=12.9$) as
\cite{Praszalowicz:2004dn}:  
\begin{equation}
\alpha=336.4-12.9\,\Sigma_{\pi N},\quad\beta=-336.4+4.3\,\Sigma_{\pi N}
,\quad\gamma=-475.94+8.6\,\Sigma_{\pi N}  \label{albega}
\end{equation}
(in units of MeV). Moreover, the inertia parameters which describe the
representation splittings take the following values (in MeV) 
\begin{equation}
\frac{1}{I_{1}}=152.4,\quad\frac{1}{I_{2}}=608.7-2.9\,\Sigma_{\pi N}.
\label{ISigma}
\end{equation}
Equations (\ref{albega}) and (\ref{ISigma}) follow from the fit to the
masses of octet and decuplet baryons and of $\Theta^{+}$ as well. If,
furthermore, one imposes the additional constraint that $M_{\Xi_{\overline{10%
}}}=1860$ MeV, then $\Sigma_{\pi N}=73$ MeV \cite{Praszalowicz:2004dn} (see
also \cite{Schweitzer:2003fg}) in agreement with recent experimental
estimates~\cite{Sigma}.

Because the Hamiltonian of Eq.(\ref{Hsplit}) mixes different SU(3)
representations, the collective wave functions are given as linear
combinations \cite{Kim:1998gt}: 
\begin{eqnarray}  \label{admix}
\left\vert B_{8}\right\rangle &=&\left\vert 8_{1/2},B\right\rangle +c_{%
\overline{10}}^{B}\left\vert \overline{10}_{1/2},B\right\rangle
+c_{27}^{B}\left\vert 27_{1/2},B\right\rangle ,\cr \left\vert
B_{10}\right\rangle &=&\left\vert 10_{3/2},B\right\rangle
+a_{27}^{B}\left\vert 27_{3/2},B\right\rangle +a_{35}^{B}\left\vert
35_{3/2},B\right\rangle ,\cr \left\vert B_{\overline{10}}\right\rangle
&=&\left\vert \overline{10} _{1/2},B\right\rangle +d_{8}^{B}\left\vert
8_{1/2},B\right\rangle +d_{27} ^{B}\left\vert 27_{1/2},B\right\rangle +d_{%
\overline{35}}^{B}\left\vert \overline{35}_{1/2},B\right\rangle,
\end{eqnarray}
where $\left\vert B_{\mathcal{R}}\right\rangle $ denotes the state which
reduces to the SU(3) representation $\mathcal{R}$ in the formal limit $%
m_{s}\rightarrow0$. The spin index $J_{3}$ has been suppressed. The $m_{s}$%
-dependent (through the linear $m_{s}$ dependence of $\alpha$, $\beta$ and $%
\gamma$) coefficients in Eq.(\ref{admix}) can be found in Ref.\cite{Yang}.

The magnetic moment collective operator can be parameterized by six
constants that in the \emph{model independent approach} are treated as free~%
\cite{Kim:1997ip,Kim:1998gt}: 
\begin{eqnarray}  \label{eq:colmag}
\hat{\mu}^{(0)} &=&w_{1}D_{Q3}^{(8)}\;+\;w_{2}d_{pq3}D_{Qp}^{(8)}\cdot 
\hat{J}_{q}\;+\;\frac{w_{3}}{\sqrt{3}}D_{Q8}^{(8)}\hat{J}_{3},\cr \hat{\mu}%
^{(1)} &=&\frac{w_{4}}{\sqrt{3}}d_{pq3}D_{Qp}^{(8)}D_{8q}
^{(8)}+w_{5}\left(D_{Q3}^{(8)}D_{88}^{(8)}+D_{Q8}^{(8)}D_{83}^{(8)}\right)
\;+\;w_{6}\left(D_{Q3}^{(8)}D_{88}^{(8)}-D_{Q8}^{(8)}D_{83}^{(8)}\right).
\end{eqnarray}
Parameters $w_{4,5,6}$ are of the order $\mathcal{O}(m_{s})$.

The full expression for the magnetic moments can be decomposed as follows%
\begin{equation}
\mu_{B}=\mu_{B}^{(0)}+\mu_{B}^{(\mathrm{op})}+\mu_{B}^{(\mathrm{wf})}
\end{equation}
where $\mu_{B}^{(0)}$ is given by the matrix element of $\hat{\mu}^{(0)}$
between the pure symmetry states $\left\vert \mathcal{R}_{J},B,J_{3}\right%
\rangle $, and $\mu_{B}^{(\mathrm{op})}$ is given as the matrix element of $%
\hat{\mu}^{(1)}$ between the symmetry states as well. Wave function
correction $\mu_{B}^{(\mathrm{wf})}$ is given as a sum of the interference
matrix elements of $\mu_{B}^{(0)}$ between pure symmetry states and
admixtures displayed in Eq.(\ref{admix}). These matrix elements were
calculated for octet and decuplet baryons in Ref.\cite{Kim:1998gt} and for
antidecuplet in Ref.\cite{Yang}.

In Ref.\cite{Yang} an overal fit to octet magnetic moments has been
performed with input parameters given by Eq.(\ref{albega}). The resulting
values for constants $w_{i}$ can be conveniently parameterized as follows: 
\begin{eqnarray}  \label{wis}
w_{1} &=&-3.7357-0.1073\,\Sigma_{\pi N},  \notag \\
w_{2} &=&24.3698-0.2146\,\,\Sigma_{\pi N},  \notag \\
w_{3} &=&7.547,  \notag \\
w_{4} &=&-5.1642-0.1332\,\,\Sigma_{\pi N}-(0.0304\,\,\Sigma_{\pi
N})^{2}+(0.0321\,\,\Sigma_{\pi N})^{3},  \notag \\
w_{5} &=&-3.742,  \notag \\
w_{6} &=&-2.443.
\end{eqnarray}
Interestingly only $w_{4}$ is a nonlinear function of $\Sigma_{\pi N}$. The
fit presented in Eq.(\ref{wis}) works well for $\Sigma_{\pi N}=45\sim 75$
MeV. 

\section{Transition magnetic moments}


\subsection{General formalism}


The transition form factors $F_i(q^2)$ from the baryon antidecuplet to the
octet are expressed in the quark matrix elements as follows: 
\begin{equation}
\langle B_8(p^{\prime}) | J_\mu (0) | B_{\overline{10}} (p)\rangle = \bar{u}%
_{B_8} ({\bm p}^{\prime},\lambda^{\prime}) \left[F_1 (q^2) +
i\sigma_{\mu\nu} \frac{q^\nu}{M_8+M_{\overline{10}}} F_2 \right] u_{B_{%
\overline{10}}}({\bm p}, \lambda),  \label{eq:em1}
\end{equation}
where $q^2$ is the square of the four-momentum transfer, and $M_8$ and $M_{%
\overline{10}}$ denote the mass of the baryon octet and antideucplet,
respectively. $u_{B_i}$ are the corresponding spinors. The electromagnetic
quark current $J_\mu$ is defined as 
\begin{equation}
J_\mu (x) = \bar{\psi}(x) \gamma_\mu \hat{Q} \psi(x)
\end{equation}
with the charge operator of the quark field $\psi$: 
\begin{equation}
\hat{Q} = \left( 
\begin{array}{ccc}
\frac23 & 0 & 0 \\ 
0 & -\frac13 & 0 \\ 
0 & 0 & -\frac13%
\end{array}
\right) = T_3 + \frac{Y}{2}.
\end{equation}
$T_3$ and $Y$ are respectively the third component of the isospin and
hypercharge given by the Gell-Mann--Nishjima formula.

Similarly, the transition form factors from the baryon decuplet to the octet
are defined as~\cite{Jones:1972ky} 
\begin{eqnarray}
\langle B_{8}(p^{\prime}) |J_{\mu} |B_{10} (p)\rangle &=& i \sqrt{\frac{2}{3}%
} \bar{u}_{B_8} ({\bm p}^{\prime},\lambda^{\prime}) \left[G_M^*(q^2)\mathcal{%
K}_{\beta\mu}^M + G_E^* (q^2) \mathcal{K}_{\beta\mu}^E\right.\cr &&\left. 
\hspace{4.5cm} + G_C^* (q^2) \mathcal{K}_{\beta\mu}^C\right]
u_{B_{10}}^{\beta} ({\bm p}, \Lambda),  \label{eq:em2}
\end{eqnarray}
where $G_M^*$, $G_E^*$, and $G_C^*$ are, respectively, the magnetic dipole,
electric quadrupole, and Coulomb form factors. The $\mathcal{\ K}^{M,E,C}$
stand for the corresponding covariant tensors~\cite{Jones:1972ky}: 
\begin{eqnarray}
\mathcal{K}_{\beta\mu}^{M} &=& -i\frac{3(M_{10}+M_8)}{2M_8
[(M_{10}+M_8)^2-q^2]} \epsilon_{\beta\mu\lambda\sigma}P^\lambda q^\sigma ,%
\cr {\cal K}_{\beta\mu}^{E} &=& -\mathcal{K}_{\beta\mu}^M - \frac{%
6(M_{10}+M_8)}{M_8\Delta(q^2)}\epsilon_{\beta\sigma\lambda\rho} P^\lambda
q^\rho \epsilon^{\sigma}_{\mu\kappa\delta} P^\kappa q^\delta \gamma^5 ,\cr
{\cal K}_{\beta\mu}^C &=& -i\frac{3(M_{10}+M_8)}{M_8 \Delta(q^2)}
q_\beta(q^2 P_\mu - q\cdot Pq_\mu)\gamma^5
\end{eqnarray}
with 
\begin{equation}
\Delta(q^2) = [(M_{10}+M_8)^2 - q^2][(M_{10}-M_8)^2-q^2].
\end{equation}
At $q^2=0$ the transition magnetic dipole form factors $F_2(q^2)$ and $%
G_M^*(q^2)$ are identified as the transition magnetic moments.

Since the magnetic dipole transitions ($M1$) are experimentally dominant
over the electric quadrupole transitions ($E2$) in hyperon radiative decays,
one can neglect the $E2$ transitions. Thus, we can express the partial decay
width in terms of the transition magnetic moments. Using Eqs.~(\ref{eq:em1},%
\ref{eq:em2}) and neglecting the $E2$ transitions, we obtain the partial
width of radiative decays from the baryon antidecuplet to the octet and from
the decuplet to the octet, respectively: 
\begin{eqnarray}  \label{eq:parwidth}
\Gamma(B_{\overline{10}}\to B_8\gamma) &=& 4 \alpha_{\mathrm{EM}} \frac{%
E_\gamma^3 }{(M_8 + M_{\overline{10}})^2} \left(\frac{{\mu}_{B_8B_{\overline{%
10}}}}{\mu_N}\right)^2,\cr \Gamma(B_{10} \to B_8\gamma) &=& \frac{\alpha_{%
\mathrm{EM}}}{2} \frac{E_\gamma^3}{M_8^2} \left(\frac{\mu_{B_8B_{10}}}{\mu_N}%
\right)^2,
\end{eqnarray}
where $\alpha_{\mathrm{EM}}$ denotes the fine structure constant and $%
E_\gamma$ is the energy of the produced photon: 
\begin{equation}
E_\gamma = \frac{M_{10 (\overline{10})}^2-M_8^2}{2M_{10 (\overline{10})}}.
\end{equation}

In the present work, we are interested in the following transition magnetic
moments (in units of nuclear magneton $\mu_{N}$): 
\begin{eqnarray}  \label{mutransdef}
\mu_{N\Delta} &=& \left\langle N,\frac{1}{2}\right\vert \hat{\mu}\left\vert
\Delta,\frac{1}{2}\right\rangle,\; \;\mu_{\Lambda^0\Sigma^0}=\left\langle
\Lambda^0,\frac{1}{2}\right\vert \hat{\mu}\left\vert \Sigma^0,\frac{1} {2}%
\right\rangle, \cr \mu_{\Sigma\Sigma^{\ast}} &=& \left\langle\Sigma,\frac{1}{%
2}\right\vert \hat{\mu}\left\vert \Sigma^{\ast},\frac{1}{2}%
\right\rangle,\;\; \mu_{\Lambda^0\Sigma^{\ast 0}}=\left\langle \Lambda^0,%
\frac{1}{2}\right\vert \hat{\mu}\left\vert \Sigma^{\ast 0},\frac{1} {2}%
\right\rangle, \cr \mu_{\Xi\Xi^{\ast}} &=& \left\langle\Xi,\frac{1}{2}%
\right\vert \hat{\mu}\left\vert \Xi^{\ast},\frac{1}{2}\right\rangle,\;\;
\mu_{NN_{\overline{10}}^{\ast}}=\left\langle N,\frac{1}{2}\right\vert \hat{%
\mu}\left\vert N_{\overline{10}}^*,\frac{1}{2}\right\rangle .
\end{eqnarray}

\subsection{Analysis of experiments}

The experimental value~\cite{PDG2004} is known: 
\begin{equation}
|\mu_{\Lambda^0\Sigma^0}|= (1.61\pm0.08)\,\mu_N.
\end{equation}

The partial decay widths for $\Delta ^{+}\rightarrow p+\gamma $ are
expressed in terms of helicity amplitudes $A_{3/2}$ and $A_{1/2}$ for
radiative decays~\cite{PDG2004}:%
\begin{eqnarray}
\Gamma (\Delta \rightarrow p\gamma )&=&\frac{E_{\gamma }^{2}}{4\pi
}\frac{M_{8}}{M_{10}}\frac{8}{2J+1}\left[ 
|A_{1/2}|^{2}+|A_{3/2}|^{2}\right] \\
&=&\frac{E_{\gamma }^{2}}{4\pi
}\frac{M_{8}}{M_{10}}\frac{8}{2J+1}\left[
  |M_{1}|^{2}+3|E_{2}|^{2}\right]  
\label{eq:decay0}
\end{eqnarray}
where 
\begin{equation}
A_{1/2}=-\frac{1}{2}(M_{1}+3E_{2}),\;\;A_{3/2}=-\frac{\sqrt{3}}{2}
(M_{1}-E_{2})
\end{equation}
with the magnetic dipole ($M_{1}$) and electric quadrupole ($E_{2}$)
amplitudes. Eq.(\ref{eq:decay0}) being used, the empirical value of the $\mu
_{p\Delta ^{+}}$ can be extracted from the data for the helicity amplitudes $
\Delta ^{+}\rightarrow p+\gamma $~\cite{PDG2004}: 
\begin{eqnarray}
A_{1/2} &=&(-0.135\pm 0.006)\,\mathrm{GeV}^{-1/2},  \notag \\
A_{3/2} &=&(-0.250\pm 0.008)\,\mathrm{GeV}^{-1/2}
\end{eqnarray}%
and the ratio $E_{2}/M_{1}$~\cite{PDG2004}: 
\begin{equation}
\frac{E_{2}}{M_{1}}=-0.025\pm 0.005.
\end{equation}
The extracted value for the $p\Delta ^{+}$ transition magnetic moment is
approximately: 
\begin{equation}
|\mu _{p\Delta ^{+}}|\simeq 3.1\,\mu _{N}.
\end{equation}
Note that the $M_{1}$ transition amplitude~\cite{Sato:2000jf} is related to
the transition magnetic moment by 
\begin{equation}
M_{1}=\frac{e}{2M_{8}}\sqrt{\frac{M_{10}E_{\gamma }}{M_{8}}}
\left( \frac{\mu _{B_{8}B_{10}}}{\mu _{N}}\right) ,  \label{eq:m1}
\end{equation}
where $e$ is the electric charge. Putting Eq.(\ref{eq:m1}) into Eq.(\ref%
{eq:decay0}), we eaily find that it is just the same as Eq.(\ref{eq:parwidth}%
).

Recently, the SELEX collaboration~\cite{Molchanov:2004iq} has announced the
upper limit on the partial width for the radiative decay of $\Sigma^{\ast-}
(1385)$ $(J^P = \frac{3}{2}^+)$: 
\begin{equation}
\Gamma (\Sigma^{\ast-}\rightarrow \Sigma^-\gamma) < 9.5\,\mathrm{keV}
\end{equation}
at $90\%$ confidence level. Using Eq.(\ref{eq:parwidth}), we can extract the
upper bound for the $\Sigma^{\ast-}\Sigma^-$ transition magnetic moment from
the SELEX data: 
\begin{equation}
\left\vert \mu_{\Sigma^{-}\Sigma^{\ast-}}\right\vert <0.82 \,\mu_N.
\label{eq:selex0}
\end{equation}
Since the radiative decay $\Sigma^{\ast -}\rightarrow \Sigma^- + \gamma$ is
forbidden in SU(3) symmetry~\cite{Polyakov:2003dx,Beg:1964nm}, the pertinent
magnetic transition moment provides us the measure of explicit
SU(3)-symmetry breaking.

The GRAAL experiment~\cite{GRAAL} indicated that a possible resonant
structure with a narrow peak at $W=1.67$ GeV could exist in $\eta $
photoproduction off the neutron, i.e. $\gamma n\rightarrow \eta n$, which was
not found in $\gamma p\rightarrow \eta p$ reaction.  This resonant peak looks
promising as a candidate for the non-strange pentaquark baryon, though one
should not exclude that it could be a manifestation of one of known
resonances such as $D_{15}(1675)$~\cite{GRAAL}.  If the observed peak
originates from the excitation of the pentaquark baryon
$n_{\overline{10}}^{\ast }$, it is of great importance to investigate
the transition magnetic moments between the non-strange pentaquarks
and the nucleons, since their partial decay widths are proportional to
$|\mu _{N_{\overline{10}}^{\ast }N}|^{2}$.  Indeed, we will show that
the present results are compatible with the GRAAL experiment, assuming
that the observed peak is in fact interpreted as the non-strange
pentaquark $n_{\overline{10}}^{\ast }$. 

\subsection{Parametrization of transition magnetic moments}

The predictions for the matrix elements entering Eqs.(\ref{mutransdef}) in
the $\chi$QSM are given below.  As for the leading order ($m_{s}=0$),
we obtain the following expressions:  
\begin{eqnarray}  \label{mu0}
\mu_{N\Delta}^{(0)}
&=&\frac{1}{3\sqrt{5}}(w_{1}-\frac{1}{2}w_{2}),\hspace{1.98cm}
\mu_{\Lambda^0\Sigma^0}^{(0)} =-\frac{\sqrt{3}}{20}(w_{1}-\frac{1}{2} 
w_{2}+\frac{1}{6}w_{3}),\cr \mu_{\Sigma\Sigma^{\ast}}^{(0)} &=&-\frac{1}{6
\sqrt{5}}(Q+1)(w_{1}-\frac {1}{2}w_{2}),\hspace{0.3cm} \mu_{\Lambda^0
\Sigma^{\ast 0}}^{(0)} =\frac{1}{2\sqrt{15}}(w_{1}-\frac{1}{2}w_{2}),\cr 
\mu_{\Xi\Xi^{\ast}}^{(0)} &=&-\frac{1}{3\sqrt{5}}(Q+1)(w_{1}-\frac {1}{2}
w_{2}),\;\; \mu_{NN_{\overline{10}}^{\ast}}^{(0)} = -\frac{1}{6\sqrt{5}}
(Q-1)(w_{1}+w_{2}+\frac{1}{2}w_{3}).
\end{eqnarray}
Let us note that both $\mu_{\Sigma^{-}\Sigma^{\ast-}}^{(0)}$ and
$\mu_{pp_{\overline{10}}^{\ast}}^{(0)}$ vanish in this order due to
the charge factors $Q=\pm1$, respectively.  As mentioned above, this is
entirely due to the SU(3) flavor symmetry: The $\chi$QSM provides a
link between various reduced matrix elements expressed in terms of the
constants $w_{1,2,3}$. Note that the transitions from the nonexotic
baryon decuplet and the octet states depend only on the combination
$w_{1}-w_{2}/2$ and therefore $w_{1}$ and $w_{2}$ cannot be extracted
separately in the zeroth order in $m_{s}$. The deviation from zero of
$\mu_{\Sigma^{-}\Sigma^{\ast-}}^{(0)}$,
$\mu_{\Xi^{-}\Xi^{\ast-}}^{(0)}$, and
$\mu_{pp_{\overline{10}}^{\ast}}^{(0)}$ is entirely due to the
symmetry breaking terms. Thus, these transition magnetic moments will
measure directly the strength of SU(3)-symmetry breaking in light
baryons. Let us also note that the $\Delta$-N transitions do not
depend on charge, which will be also true for their $m_s$
corrections. 

We find that the transition magnetic moments of Eq.(\ref{mu0}) satisfy the
following relations based on pure SU(3)-symmetry: 
\begin{eqnarray}  \label{eq:symm}
\mu_{N\Delta} &=& -\mu_{\Sigma^+\Sigma^{\ast +}} =
-2\mu_{\Sigma^0\Sigma^{\ast 0}} = \frac{\sqrt{2}}{3} \mu_{\Lambda^0\Sigma^{
\ast 0}} = -\mu_{\Xi^0\Xi^{\ast 0}},\cr \mu_{\Sigma^-\Sigma^{\ast-}} &=&
\mu_{\Xi^-\Xi^{\ast-}}=0.
\end{eqnarray}
which are the same as those in Ref.~\cite{Beg:1964nm} apart from the
relative signs. The sign difference arises from the different convention for
the baryon states. In addition, the transition magnetic moment
$\mu_{\Lambda^0\Sigma^0}$ can be expressed in terms of
$\mu_{\Lambda^0}$ and $\mu_{\Sigma^0}$: 
\begin{equation}
\mu_{\Lambda^0\Sigma^0} = \frac{\sqrt{3}}{2} (\mu_{\Sigma^0} -
\mu_{\Lambda^0}).
\end{equation}

The wave function corrections read as follows: 
\begin{eqnarray}  \label{eq:wf}
\mu_{N\Delta}^{(\mathrm{wf})} & = & \frac{1}{9\sqrt{5}}c_{27}(w_{1}+2w_{2})+%
\frac {5}{9\sqrt{5}}a_{27}\left(w_{1}+\frac{1}{2}w_{2}\right),\cr %
\mu_{\Lambda\Sigma}^{(\mathrm{wf})} &=& -\frac{1}{4\sqrt{3}}c_{\overline{10}%
} \left(w_{1}+w_{2}+\frac{1}{2}w_{3}\right) -\frac{1}{12\sqrt{3}}%
c_{27}\left(w_{1}+2w_{2}-\frac{3}{2}w_{3}\right),\cr \mu_{\Sigma\Sigma^{%
\ast}}^{(\mathrm{wf})} & = &\frac{1}{18\sqrt{5}}c_{27}
(2-3Q)\left(w_{1}+2w_{2}\right) +\frac{2}{9\sqrt{5}}a_{27}\left(w_{1}+\frac{1%
}{2}w_{2}\right),\cr \mu_{\Lambda\Sigma^{*0}}^{(\mathrm{wf})} &=& \frac{1}{3%
\sqrt{15}} c_{27} \left(w_{1}+2 w_{2} \right) + \frac{2}{3\sqrt{15}}
a_{27}\left(w_{1}+\frac12 w_{2}\right),\cr \mu_{\Xi_{8}\Xi_{10}^{*}}^{(%
\mathrm{wf})}&=&-\frac{1}{18\sqrt{15}} c_{27}(7Q+2)(w_{1}+2w_{2})
\,-\,\frac1{9\sqrt{5}}a_{27} \,(Q-1)\left(w_{1}+\frac12w_{2}\right)\,,\cr %
\mu_{NN_{\overline{10}}^{\ast}}^{(\mathrm{wf})} &=&\frac{1}{\sqrt{5}}c_{%
\overline {10}}\left[ \frac{5}{48}Q\,\left(w_{1}-\frac{7}{2}w_{2}-\frac{1}{2}%
w_{3}\right) +\frac{1}{8}(6Q-2)\left(w_{1}-\frac{1}{2}w_{2}+\frac{1} {6}%
w_{3}\right)\right] \cr && -\frac{1}{180\sqrt{5}}d_{27}(Q+2)%
\left(w_{1}+2w_{2}-\frac{3}{2} w_{3}\right)\cr &&-\frac{7}{144\sqrt{5}}%
c_{27}(7Q-4)\left(w_{1}-\frac{11}{14}w_{2} - \frac{3}{14}w_{3}\right),
\end{eqnarray}
where we have used the relation $d_{8}=-c_{\overline{10}}$~\cite{Yang}.
Finally, the operator parts of the linear $m_{\mathrm{s}}$ corrections read 
\begin{eqnarray}  \label{eq:op}
\mu_{N\Delta}^{(\mathrm{op})} & =&\frac{1}{108\sqrt{5}}%
(7w_{4}+15w_{5}+9w_{6}),\cr \mu_{\Lambda\Sigma}^{(\mathrm{op})} &=&-\frac{1}{%
180\sqrt{3}}(7w_{4}+6w_{5}),\cr \mu_{\Sigma\Sigma^{\ast}}^{(\mathrm{op})} &=&%
\frac{1}{108\sqrt{5}}((4-3Q)\,w_{4}+3(2-3Q)\,w_{5}+9Q\,w_{6}),\cr %
\mu_{\Lambda\Sigma^{*0}}^{(\mathrm{op})} & = & \frac{1}{18\sqrt{15}}%
\left(2w_4 +3w_5\right),\cr \mu_{\Xi_{8}\Xi_{10}^{*}}^{(\mathrm{op})} &=& -%
\frac{1}{108\sqrt{5}}\left[(8Q+1)w_{4}+3(4Q-1)w_{5}-9(2Q+1)w_{6}\right],\cr %
\mu_{NN_{\overline{10}}^{\ast}}^{(\mathrm{op})} &=& -\frac{1}{54\sqrt{5}}
(Q+1)w_{4}-\frac{1}{18\sqrt{5}}(2Q-1)\left(w_{5}+\frac{3}{2}w_{6}\right).
\end{eqnarray}

Looking at Eqs.(\ref{mu0},\ref{eq:wf},\ref{eq:op}), we find an interesting
relation: Though the general expressions for the transitions
$\Sigma^{\ast}\rightarrow \Sigma$ and $\Xi^{\ast}\rightarrow \Xi$ look 
different, they turn out to be the same for the negative charge $Q=-1$, 
\emph{i.e.} for the transition magnetic moments $\Sigma^{\ast -}\rightarrow
\Sigma^-$ and $\Xi^{\ast -}\rightarrow \Xi^-$.


\section{Results and Discussion}

\begin{figure}[h]
\begin{center}
\includegraphics[scale=0.47]{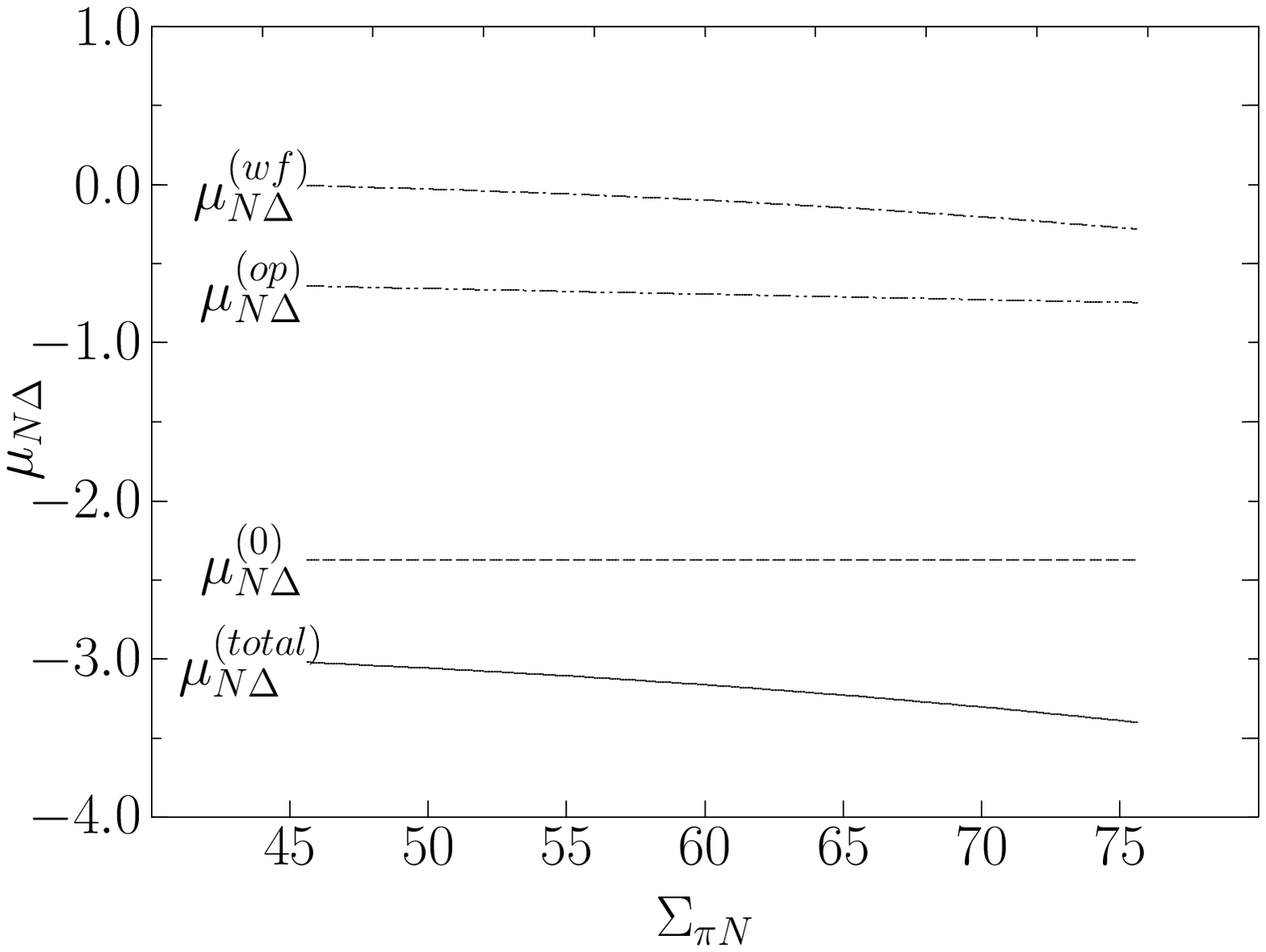} \hspace{0.5cm} %
\includegraphics[scale=0.47]{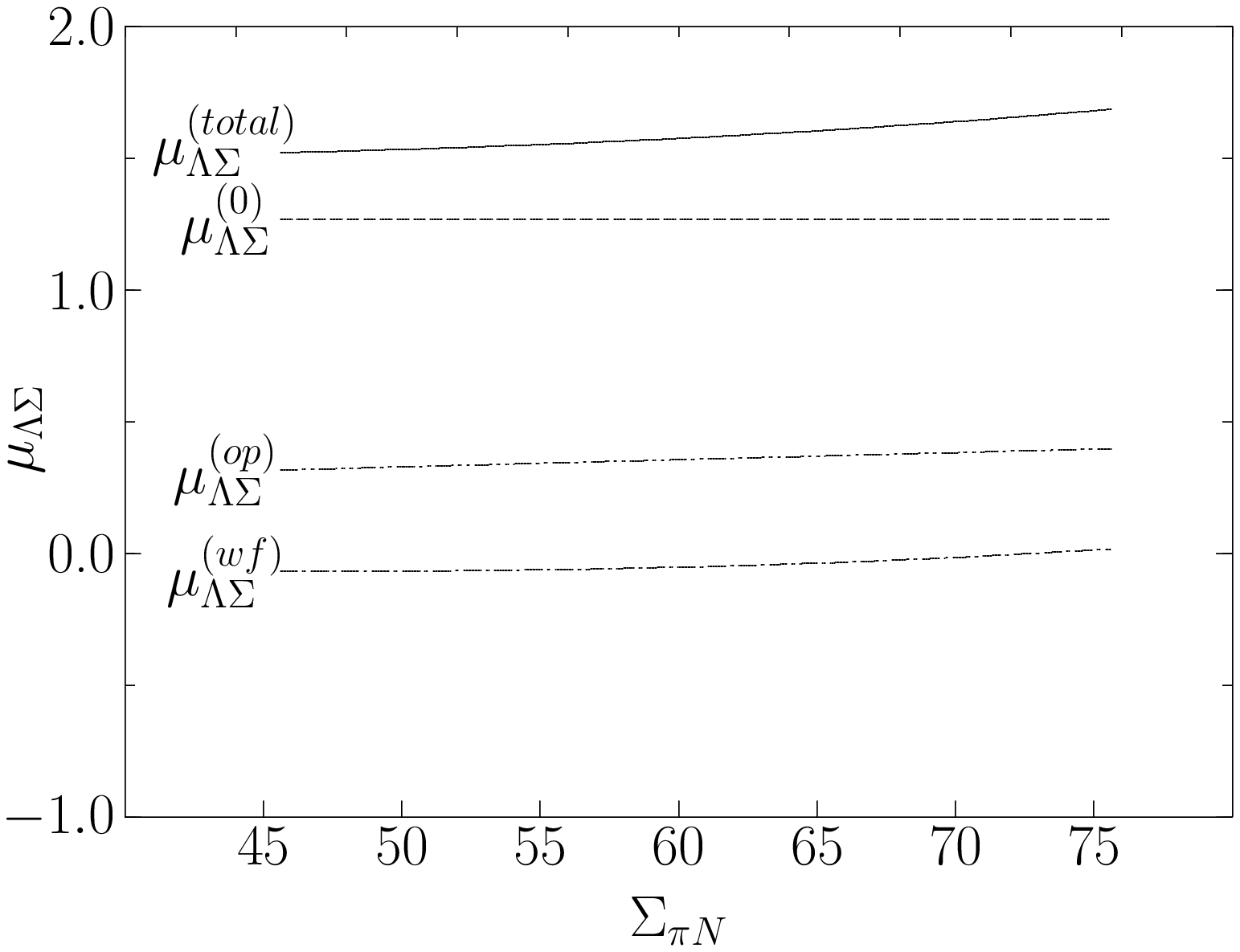}
\end{center}
\caption{$\Delta\rightarrow N$ transition magnetic moment as function of $%
\Sigma_{\protect\pi N}$ in the left panel and $\Sigma \rightarrow \Lambda$
one as function of $\Sigma_{\protect\pi N}$ in the right panel.}
\label{fig:1}
\end{figure}

Incorporating the parameterizations given in Eqs. (\ref{albega}) and (\ref%
{wis}), we obtain the numerical results for the nonexotic and exotic
transition magnetic moments, which are summarized in Figs.~\ref{fig:1}--\ref%
{fig:5}. 
\begin{figure}[h]
\begin{center}
\includegraphics[scale=0.47]{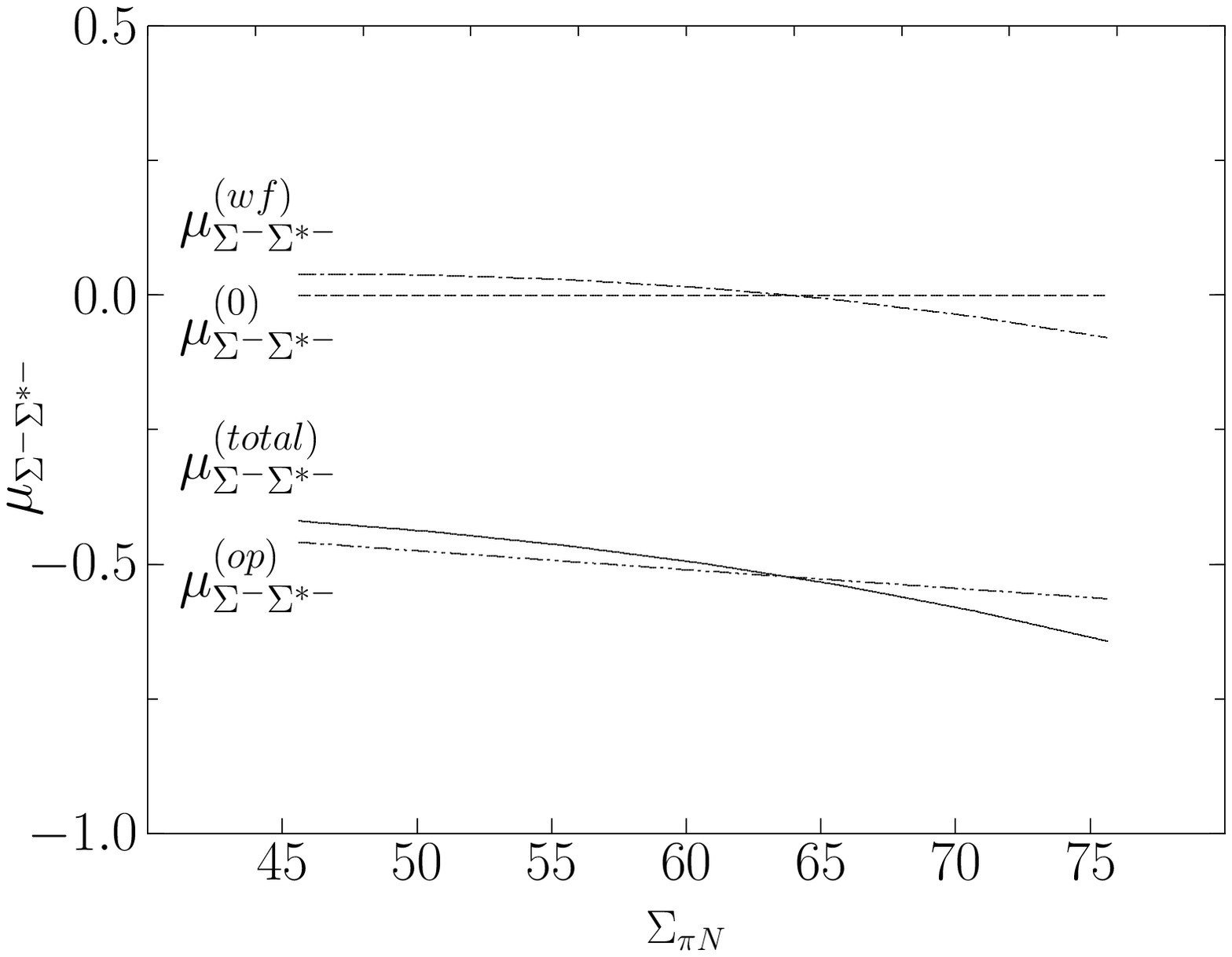} \hspace{0.5cm} %
\includegraphics[scale=0.47]{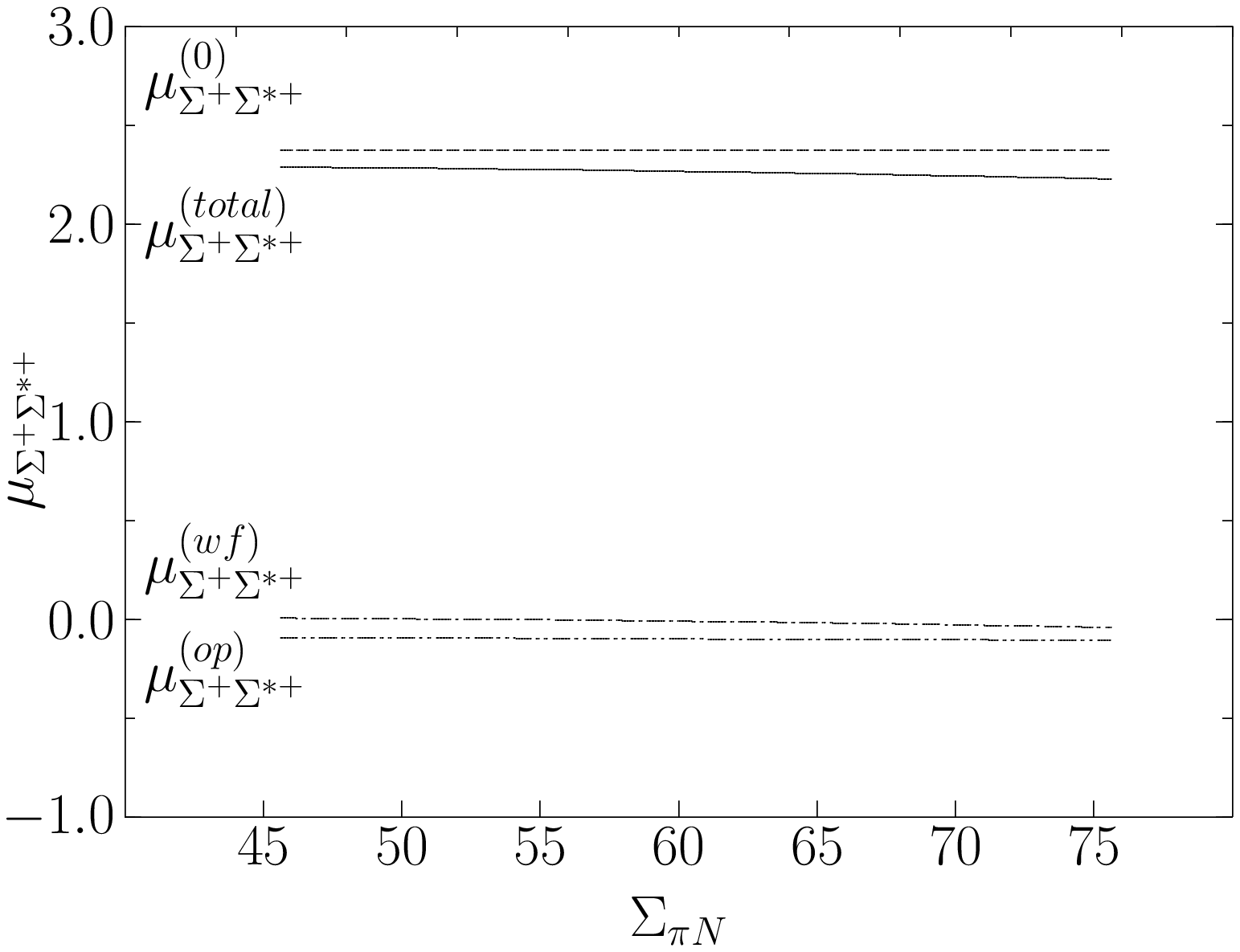}
\end{center}
\caption{$\Sigma^{\ast-}\rightarrow \Sigma^-$ transition magnetic moment as
functions of $\Sigma_{\protect\pi N}$ in the left panel and $%
\Sigma^{\ast+}\rightarrow \Sigma^+$ one as functions of $\Sigma_{\protect\pi %
N}$ in the right panel.}
\label{fig:2}
\end{figure}
\begin{figure}[h]
\begin{center}
\includegraphics[scale=0.47]{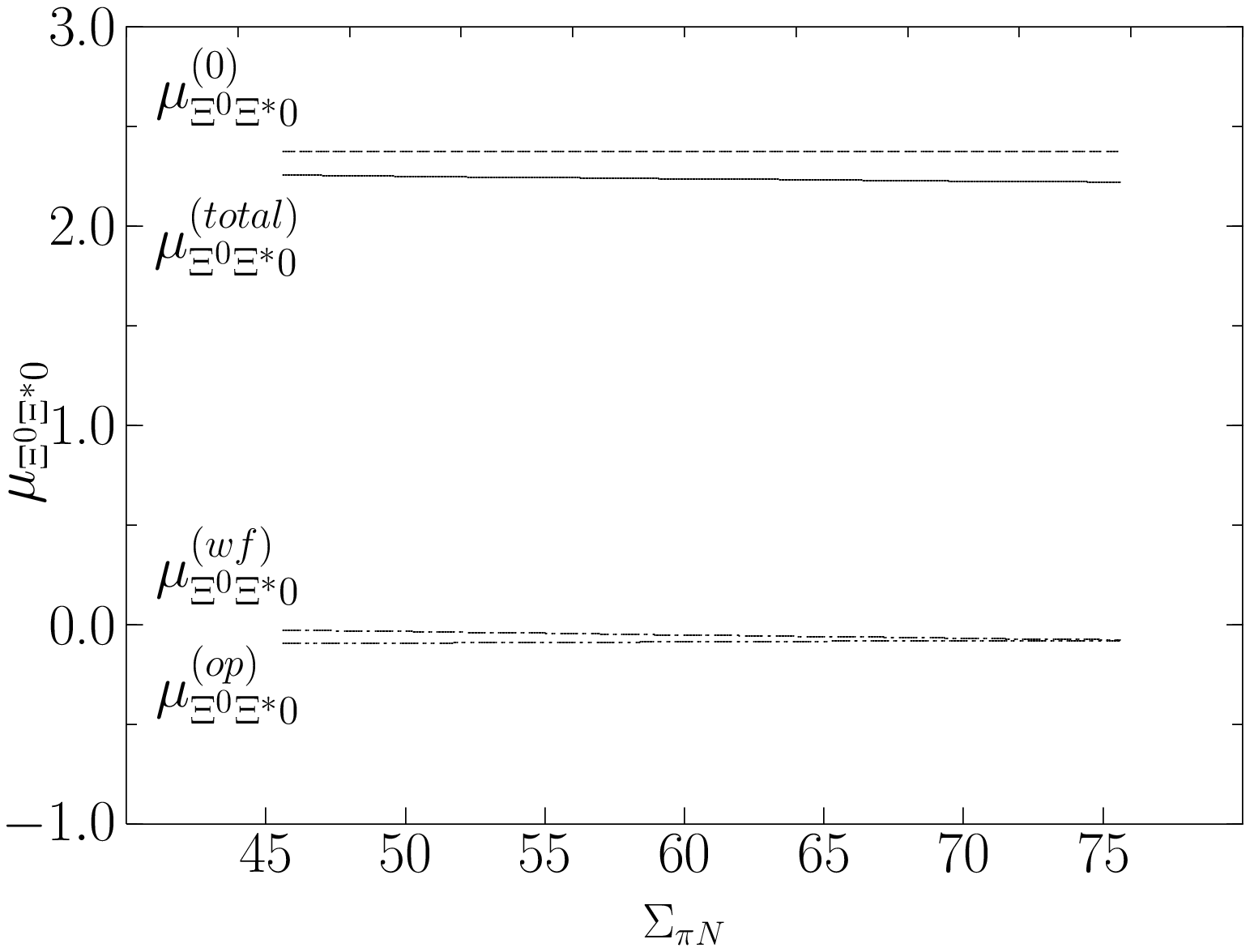} \hspace{0.5cm} %
\includegraphics[scale=0.47]{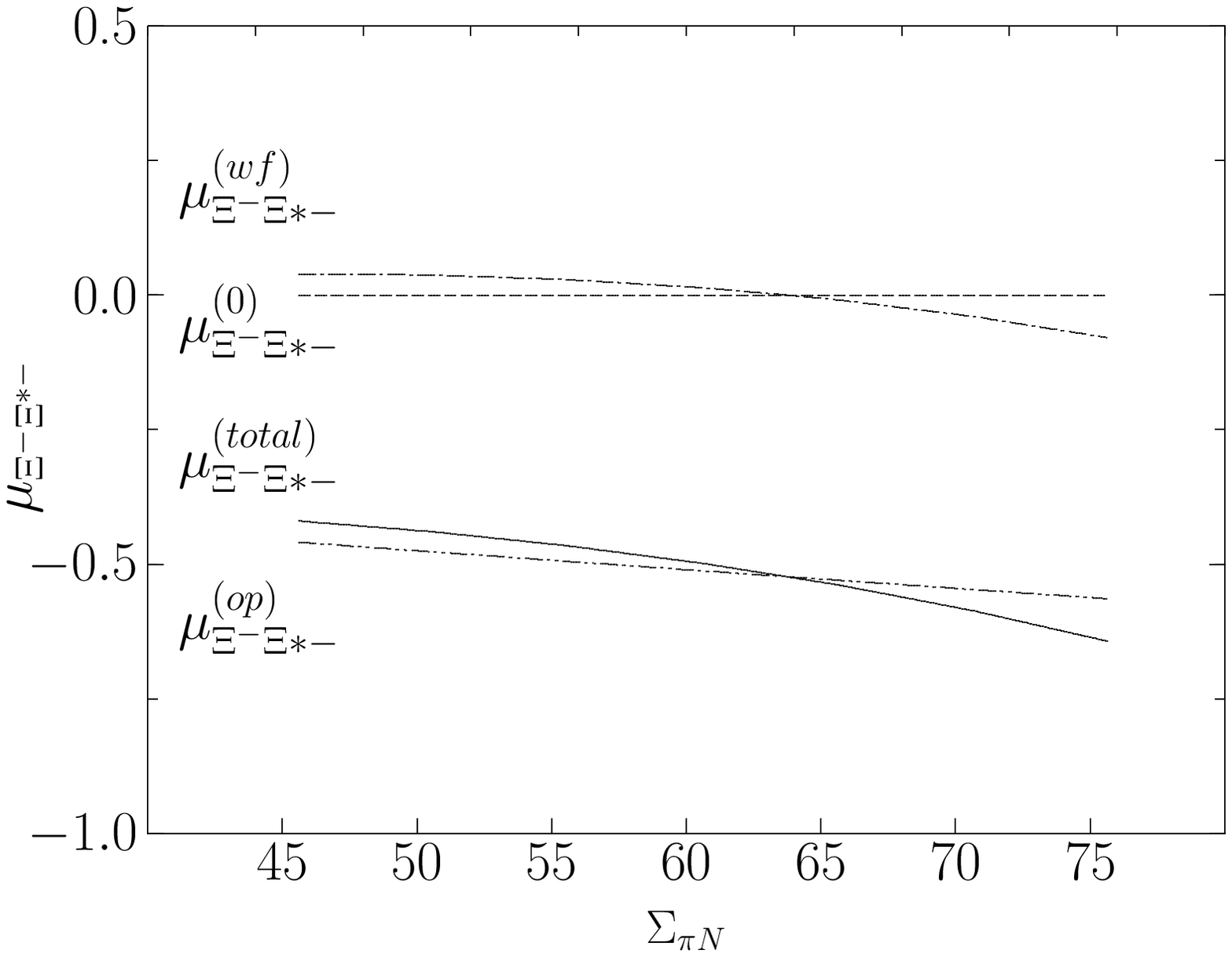}
\end{center}
\caption{$\Xi^{\ast 0 }\rightarrow \Xi^{0}$ transition magnetic moment as
functions of $\Sigma_{\protect\pi N}$ in the left panel and $\Xi^{\ast
-}\rightarrow \Xi^-$ one as functions of $\Sigma_{\protect\pi N}$ in the
right panel.}
\label{fig:3}
\end{figure}
\begin{figure}[h]
\begin{center}
\includegraphics[scale=0.5]{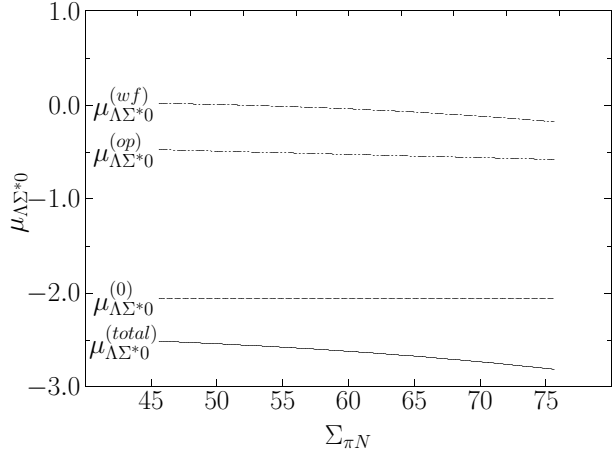}
\end{center}
\caption{$\Sigma^{\ast 0}\rightarrow \Lambda$ transition magnetic moment as functions
of $\Sigma_{\protect\pi N}$.}
\label{fig:4}
\end{figure}
\begin{figure}[h]
\begin{center}
\includegraphics[scale=0.5]{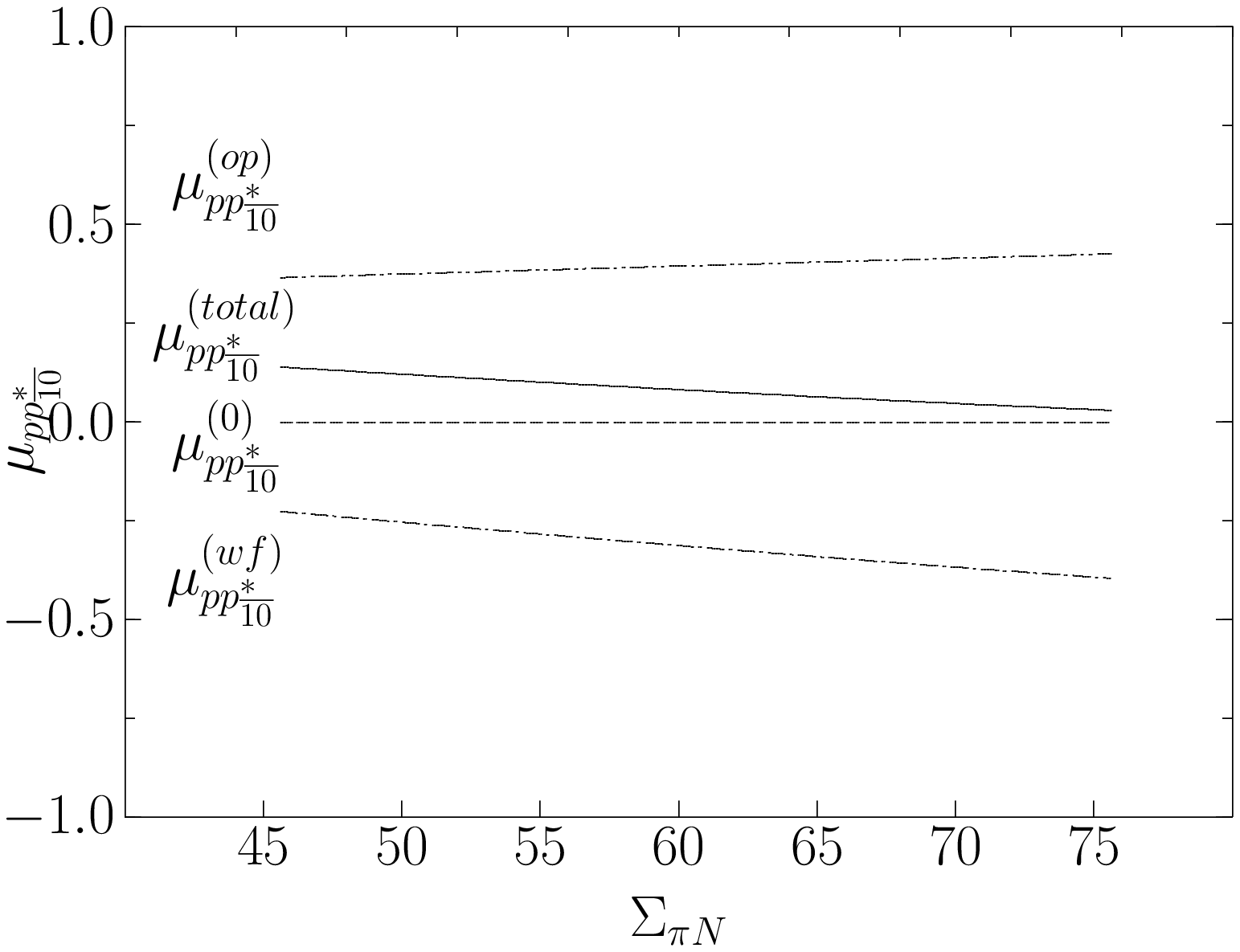} \hspace{0.5cm} %
\includegraphics[scale=0.5]{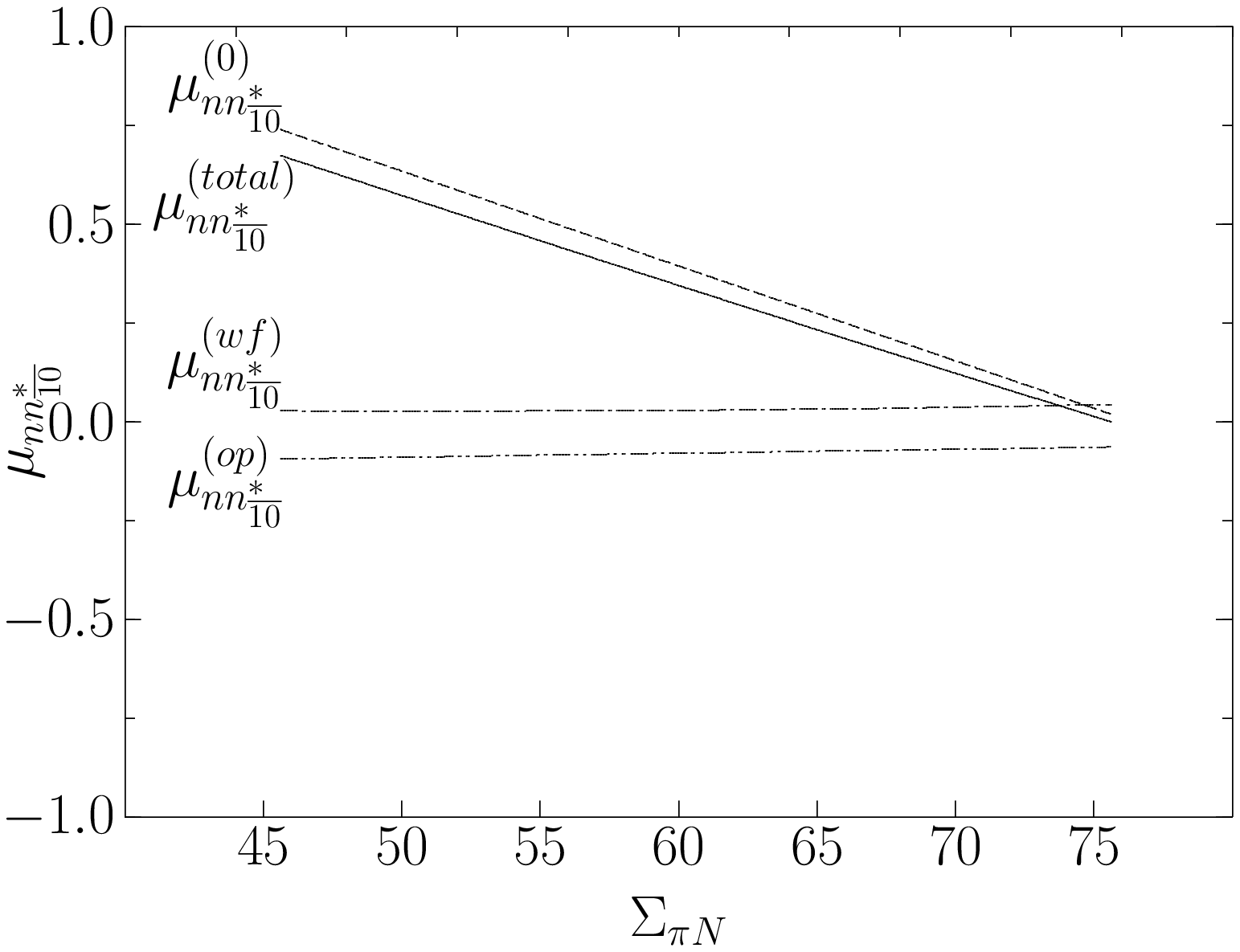}
\end{center}
\caption{$p^{\ast}\rightarrow p$ transition magnetic moment as functions of $%
\Sigma_{\protect\pi N}$ in the left panel and $n^{\ast}\rightarrow n$ one as
functions of $\Sigma_{\protect\pi N}$ in the right panel.}
\label{fig:5}
\end{figure}

It was shown in Ref.~\cite{Yang} that the magnetic moments of the baryon
decuplet are almost insensitive to the $\Sigma_{\pi N}$ term. Here we find
that almost all transition magnetic moments between the decuplet and the
octet are very weakly dependent on $\Sigma_{\pi N}$. This can be easily
understood by Eq.(\ref{mu0}): Aparently the leading-order terms are
proportional to $w_1-w_2/2$ whose dependence on $\Sigma_{\pi N}$ vanishes
according to Eq.(\ref{wis}). The $\Sigma_{\pi N}$ dependence enters only via 
$m_{\mathrm{s}}$ corrections which are rather small. The $\Sigma^{\ast
-}\rightarrow \Sigma^-$ and $\Xi^{\ast -}\rightarrow \Xi^-$ transition
magnetic moments, which have only the non-zero values with SU(3) symmetry
breaking terms, are shown to depend mildly on $\Sigma_{\pi N}$. Moreover,
these two transitions are exactly the same at each order, as depicted in
Figs~\ref{fig:2} and \ref{fig:3}.

The linear $m_{\mathrm{s}}$ corrections are almost negligible for the
transition magnetic moments $\mu_{\Sigma^+\Sigma^{\ast +}}$ and $%
\mu_{\Xi^0\Xi^{\ast 0}}$, whereas they contribute to $\mu_{N\Delta}$, $%
\mu_{\Sigma^0\Lambda^0}$, and $\mu_{\Lambda^0\Sigma^{\ast 0}}$ at around 20
\%. Since $\mu_{\Sigma^-\Sigma^{\ast -}}$ and $\mu_{\Xi^-\Xi^{\ast -}}$
vanish in pure SU(3) symmetry, their values arise soley from the effects of
the SU(3) symmetry breaking. Thus, the linear $m_{\mathrm{s}}$ terms become
a leading-order contribution in this case.

While the nonexotic magnetic transition moments are not sensitive to
$\Sigma_{\pi N}$, the exotic $\mu_{nn_{\overline{10}}^{\ast}}$ depends
rather strongly on it. It is very similar to the situation in the case
of the magnetic moments of the baryon antidecuplet~\cite{Yang}. The
reason lies in the fact that e.g. $\mu_{NN_{\overline{10}}^{\ast}}$ is
proportional to $w_{1}+w_{2}+\frac{1}{2}w_{3}$. Thus, the terms with
the $\Sigma_{\pi N}$ in $w_1$ and $w_2$ interfere constructively,
which makes $\mu_{nn_{\overline{10}}^{\ast}}$ to have approximately
linear dependence on $\Sigma_{\pi N}$, as shown in
Fig.~\ref{fig:5}. On the contrary, the transition magnetic moment
$\mu_{pp_{\overline{10}}^{\ast}}$ has non-zero value only at the
linear $m_{\mathrm{s}}$ order. Thus, its $\Sigma_{\pi N}$ dependence
arises from the $m_{\mathrm{s}}$ corrections and is rather
weak. Another interesting point is that the wave-function contribution
($\mu^{({\rm wf})}$) and the operator one ($\mu^{({\rm op})}$) almost
cancel each other, so that the linear $m_{\mathrm{s}}$ corrections for
octet-antidecuplet transitions turn out to be negligible. 

In Table~\ref{tab1} the numerical results for the transition magnetic
moments of the nonexotic and exotic baryons are listed for three different
values of the $\Sigma_{\pi N}$ in units of $\mu_N$. 
\begin{table}[h]
\begin{tabular}{c|ccccccccc}
\hline\hline
$\Sigma_{\pi N}$ [MeV] & $\mu_{N\Delta}$ & $\mu_{\Lambda^0\Sigma^0}$ & $%
\mu_{\Sigma^{-}\Sigma^{\ast-}}$ & $\mu_{\Sigma^{+}\Sigma^{\ast+}}$ & $%
\mu_{\Lambda^0\Sigma^{\ast 0}}$ & $\mu_{\Xi^0\Xi^{\ast 0}}$ & $%
\mu_{\Xi^-\Xi^{\ast -}}$ & $\mu_{pp_{\overline{10}}^{\ast}}$ & $\mu_{nn_{%
\overline{10}}^{\ast}}$ \\ \hline
$50$ & $-3.06$ & $1.54$ & $-0.44$ & $2.25$ & $-2.54$ & $2.25$ & $-0.44$ & $%
0.12$ & $0.56$ \\ 
$60$ & $-3.16$ & $1.58$ & $-0.50$ & $2.21$ & $-2.63$ & $2.24$ & $-0.50$ & $%
0.08$ & $0.33$ \\ 
$70$ & $-3.31$ & $1.64$ & $-0.59$ & $2.17$ & $-2.74$ & $2.23$ & $-0.59$ & $%
0.04$ & $0.11$ \\ \hline\hline
\end{tabular}%
\caption{Transition magnetic moments of the nonexotic and exotic baryons in
units of $\protect\mu_N$. The experimental value for $\protect\mu%
_{\Lambda^0\Sigma^0}$ is: $(1.61\pm0.08)\,\protect\mu_N$. The empirical
value for $|\protect\mu_{N\Delta}|$ is approximately equal to $3.1\,\protect%
\mu_N$.}
\label{tab1}
\end{table}
Those for $\mu_{N\Delta}$ and $\mu_{\Lambda^0\Sigma^0}$ are in a very good
agreement with the experimental data. As seen in Eq.(\ref{eq:selex0}) in the
previous section, the upper bound for $\mu_{\Sigma^{-}\Sigma^{\ast-}}$
extracted from the upper limit for the partial decay width of the SELEX
experiment is around $0.82\,\mu_N$. Compared to this, the present prediction
for $\mu_{\Sigma^{-}\Sigma^{\ast-}}$ lies definitely in the allowed region
for all reasonable values of $\Sigma_{\pi N}$. As already mentioned, the
value of $\mu_{\Xi^-\Xi^{\ast -}}$ coincides with that of $%
\mu_{\Sigma^{-}\Sigma^{\ast-}}$ even with the explicit SU(3)-symmetry
breaking considered.

\begin{table}[h]
\begin{tabular}{c|cccccccccc}
\hline\hline
$\Sigma_{\pi N}$ [MeV] & $\Gamma_{N\Delta}$ & $\Gamma_{\Lambda^0\Sigma^0}$ & 
$\Gamma_{\Sigma^{-}\Sigma^{\ast-}}$ & $\Gamma_{\Sigma^{+}\Sigma^{\ast+}}$ & $%
\Gamma_{\Lambda^0\Sigma^{\ast 0}}$ & $\Gamma_{\Xi^0\Xi^{\ast 0}}$ & $%
\Gamma_{\Xi^-\Xi^{\ast -}}$ & $\Gamma_{pp_{\overline{10}}^{\ast}}$ & $%
\Gamma_{nn_{\overline{10}}^{\ast}}$ & $\Gamma_{nn_{\overline{10}%
}^{\ast}}/\Gamma_{pp_{\overline{10}}^{\ast}}$ \\ \hline
$50$ & $672$ & $5.37$ & $2.72$ & $76.0$ & $266$ & $87.5$ & $3.05$ & $11.5$ & 
$250$ & $21.67$ \\ 
$60$ & $717$ & $5.65$ & $3.51$ & $73.3$ & $285$ & $86.8$ & $3.94$ & $5.12$ & 
$87.2$ & $17.02$ \\ 
$70$ & $786$ & $6.09$ & $4.89$ & $70.7$ & $309$ & $86.0$ & $5.48$ & $1.28$ & 
$9.69$ & $7.56$ \\ \hline\hline
\end{tabular}%
\caption{Partial decay widths for the radiative decays of exotic and
nonexotic baryons in units of keV. The last column stands for the ratio of
the partial decay widths $n_{\overline{10}}^{\ast}\rightarrow n+\protect%
\gamma$ and $p_{\overline{10}}^{\ast}\rightarrow p+\protect\gamma$.}
\label{tab2}
\end{table}
In Table~\ref{tab2} we list the numerical values of the partial decay widths
for the radiative decays of exotic and nonexotic baryons in units of keV. In
the last column of Table~\ref{tab2} the ratio of the partial decay widths $%
n_{\overline{10}}^{\ast}\rightarrow n+\gamma$ and $p_{\overline{10}%
}^{\ast}\rightarrow p+\gamma$ is given. Note that this sort of partial decay
width is proportional to $|\mu_{NN_{\overline{10}}}^*|^2$. The ratio is
consistent with the finding of the GRAAL experiment~\cite{GRAAL} which has
seen the narrow peak around $1.67$ GeV in $\eta$ photoproduction off the
neutron but not for the proton target. If one assumes that the decays of $p_{%
\overline{10}}^{\ast}\rightarrow p\eta$ and $n_{\overline{10}%
}^{\ast}\rightarrow n\eta$ are identical, it indicates that the transition
magnetic moment $\mu_{_{nn_{\overline{10}}^{\ast}}}$ must be sizeably larger
than $\mu_{_{pp_{\overline{10}}^{\ast}}}$. Indeed, the present work predicts
that the partial width for the radiative decay $n_{\overline{10}%
}^*\rightarrow n+\gamma$ is $8\sim 22$ times larger than that for the $p_{%
\overline{10}}^*\rightarrow p+\gamma$. Note that a recent work~\cite%
{Hong:2004du} in a diquark picture draws a similar conclusion, though it
predicts that the partial decay width for the neutron channel is just four
times larger than that for the proton one.


\section{Conclusion and summary}


In the present work, we have investigated the transition magnetic moments
from the baryon octet to the decuplet and from the proton and neutron to the
pentaquark nucleons of the antidecuplet. We used the the \emph{%
model-independent approach} within the framework of the chiral quark-soliton
model, thereby taking explicit SU(3)-symmetry breaking into account. The
parameters of the approach are fixed by the octet magnetic moments, octet
masses, and the mass of the $\Theta^+$, where the residual freedom is
parametrized by the pion-nucleon sigma term, $\Sigma_{\pi N}$. The results
for $\mu_{N\Delta}$ and $\mu_{\Lambda^0\Sigma^0}$ are in good agreement with
the experimental and empirical data. The transition magnetic moment $%
\mu_{\Sigma^-\Sigma^{\ast -}}$, which has only a non-zero value due to
explicit SU(3)-symmetry breaking, is found to be below its upper bound
extracted from the SELEX data~\cite{Molchanov:2004iq}. We predicted also the
value of $\mu_{\Xi^-\Xi^{\ast -}}$ which turned out to be the same as that
of $\mu_{\Sigma^-\Sigma^{\ast -}}$. The measurement of these magnetic
transitions can be possibly done within the SELEX program.

The transition magnetic moment $\mu_{nn_{\overline{10}}^*}$ turns out to be
rather sensitive to the value of $\Sigma_{\pi N}$ due to the constructive
interference of the parameters $w_1(\Sigma_{\pi N})$ and $w_1(\Sigma_{\pi
N}) $. The $\mu_{pp_{\overline{10}}^*}$ has a non-vanishing value only due
to the explicit SU(3)-symmetry breaking, so that its value becomes very
small in comparison with that of the $\mu_{nn_{\overline{10}}^*}$. As a
result, the present predictions for the transition magnetic moments $%
\mu_{pp_{\overline{10}}^*}$ and $\mu_{nn_{\overline{10}}^*}$ are consistent
with the recent GRAAL data on $\gamma p\rightarrow \eta p$ and $\gamma
n\rightarrow \eta n$. This supports the view that the peak seen in the GRAAL
experiment corresponds to a neutron-like pentaquark resonance of the
antidecuplet.

\section*{Acknowledgments}

The present work is supported by the Korea Research Foundation Grant:
KRF-2003-070-C00015 (H.-Ch.K.) and by the Polish State Committee for
Scientific Research under grant 2 P03B 043 24 (M.P.). The work has also been
supported by Korean (F01-2004-000-00102-0)-German (DFG) and Polish-German
(DFG) grants. The work is partially supported by the
Transregio-Sonderforschungsbereich Bonn-Bochum-Giessen as well as by the
Verbundforschung and the International Office of the Federal Ministry for
Education and Research (BMBF).

\end{document}